\documentstyle[prd,aps,preprint]{revtex}

\begin{document}
\title{Primordial Black Hole Formation in a Double Inflation Model in 
Supergravity} 
\author{M. Kawasaki}
\address{Institute for Cosmic Ray Research, University of Tokyo,
  Tanashi 188, Japan}
\author{Naoshi Sugiyama}
\address{Department of Physics, Kyoto University, Kyoto 606-01, Japan}
\author{T. Yanagida}
\address{Department of Physics,  University of
  Tokyo, Tokyo 113, Japan}
\date{\today}

\maketitle

\begin{abstract}
It has been recently pointed out that the initial value problem in 
new inflation models is naturally solved by supergravity effects if 
there exists a pre-inflation before the new inflation.  We study this 
double inflation model in details and find that density fluctuations 
on small cosmological scales are much larger than those on large 
scales due to peculiar property of the new inflation.  We show that 
this results in production of primordial black holes which have $\sim 1 
M_{\odot}$ masses in a certain parameter region of the double inflation 
model.  We stress that these black holes may be identified with MACHOs 
observed in the halo of our galaxy.
\end{abstract}



\section{Introduction}

The new inflation model~\cite{Albrecht} is the most interesting among 
various inflation models proposed so far, since its reheating 
temperature $T_R$ is naturally low to avoid the overproduction of 
gravitinos in supergravity~\cite{Ellis}.  It has been 
shown~\cite{Moroi} that the upperbounds of the reheating temperature 
$T_R$ should be less than $10^2$GeV--$10^6$GeV in gauge-mediated 
supersymmetry (SUSY)-breaking models~\cite{Dine,Intriligator,Haba} 
since the mass of gravitino $m_{3/2}$ is predicted in a range of 
$10^2$keV--1GeV~\cite{Gouvea}.  In hidden sector SUSY-breaking 
models~\cite{Nilles}, on the other hand, $m_{3/2}\simeq 100$GeV--1TeV. 
The reheating temperature is also constrained as $T_R \lesssim 
10^{6-9}$GeV~\cite{Ellis} even in this case.  These constraints on 
$T_R$ are easily satisfied in a large class of new inflation models.

The new inflation model, however, has two serious drawbacks. One is
the fine tuning problem of the initial condition. Namely, the universe 
has to have a large region over horizons at the beginning where the
inflaton field $\phi$ is smooth and its average value is very close 
to a local maximum of the potential $V(\phi)$. Since the inflaton
potential $V(\phi)$ should be very flat to satisfy  slow-roll
condition, there is no dynamical reason for the universe to choose
such a specific initial value of the $\phi$. Another problem is
related to the fact that in the new inflation model the Hubble
parameter is much smaller than the gravitational scale. Thus, the new
inflation itself does not give a full explanation for why our universe
lived for a long time~\cite{Linde-book}.

In a recent work~\cite{Izawa}, Izawa and two of us (MK and TY) have 
shown that the above serious problems are simultaneously solved if 
there existed a pre-inflation with a sufficiently large Hubble 
parameter before the new inflation.  In this double inflation model 
the inflation dynamics in a small scale region could be different from 
those in a large scale region in general.  If the horizon scale at the 
turning epoch from one to another inflation is cosmologically 
relevant, one may expect that density fluctuations on the scales 
smaller than this horizon scale are significantly larger than 
fluctuations on the larger scales due to nature of the new inflation.  
Such an over-powered density spectrum on small scales may produce 
primordial black holes.  Recent discovery of MAssive Compact Halo 
Objects (MACHOs) by gravitational lensing effects~\cite{MACHO} has 
revived the interest in the primordial black holes.  Since the 
observed masses of MACHOs, which have not been directly observed yet, 
are about $0.5 \sim 0.6 M_\odot$, MACHOs are very unlikely to be 
standard stars such as white dwarfs or red dwarfs.  In this paper we 
show that a large amount of black holes whose mass scales are about 
$1M_\odot$ are formed in a certain parameter region of our double 
inflation model.  We find that these black holes are considered as 
possible candidates for MACHOs.\footnote{
Different models for the primordial black hole formation have been studied in 
Refs.~\cite{Yokoyama} and \cite{Garcia}.}

\section{A New inflation model}

We adopt the new inflation model proposed in Ref.~\cite{Izawa}. 
The inflaton superfield $\phi(x, \theta)$ is assumed to have an $R$
charge $2/(n+1)$ so that the following tree-level superpotential is
allowed:
\begin{equation}
        W_{0}(\phi) = -\frac{g}{n+1}\phi^{n+1},
        \label{sup-pot}
\end{equation}
where $n$ is a positive integer and $g$ denotes a coupling constant of order
one. Here and hereafter, we set the gravitational scale $M\simeq
2.4\times 10^{18}$GeV equal to unity and regard it as a plausible
cutoff in supergravity. We further assume that the continuous $U(1)_R$
symmetry is dynamically broken down to a discrete $Z_{2nR}$ at a scale
$v$ generating an effective superpotential;
\begin{equation}
        W(\phi) = v^{2}\phi - \frac{g}{n+1}\phi^{n+1}.
        \label{sup-pot2}
\end{equation}
We may consider that the scale $v^2$ 
is induced by some nonperturbative dynamics as shown in 
Ref.~\cite{Izawa}.

The $R$-invariant effective K\"ahler potential is given by 
\begin{equation}
    \label{new-kpot}
    K(\phi,\chi) = |\phi|^2 +\frac{\kappa}{4}|\phi|^4 
    + \cdots ,
\end{equation}
where $\kappa$ is a constant of order one and the ellipsis denotes
higher-order terms, which we neglect in the present analysis.

The effective potential of a scalar component of the superfield
$\phi(x,\theta)$ in supergravity is
given by
\begin{equation}
        V = e^{K(\phi)}\left\{ \left(\frac{\partial^2 
        K}{\partial\phi\partial\phi^{*}}\right)^{-1}|D_{\phi}W|^{2}
        - 3 |W|^{2}\right\},
        \label{new-pot}
\end{equation}
with 
\begin{equation}
        D_{\phi}W = \frac{\partial W}{\partial \phi} 
        + \frac{\partial K}{\partial \phi}W.
        \label{DW}
\end{equation}
This potential yields a vacuum 
\begin{equation}
    \langle \phi \rangle  \simeq  
    \left(\frac{v^2}{g}\right)^{1/n}.
\end{equation}
We have negative energy as 
\begin{equation}
        \langle V \rangle \simeq -3 e^{\langle K \rangle}
        |\langle W \rangle |^{2}
        \simeq -3 \left( \frac{n}{n+1}\right)^{2}|v|^{4}
        |\langle \phi \rangle|^{2}.
        \label{vacuum}
\end{equation}
The negative vacuum energy (\ref{vacuum}) is assumed to be canceled 
out by a SUSY-breaking effect~\cite{Izawa2} which gives a positive 
contribution $\Lambda^{4}_{SUSY}$ to the vacuum energy.  Namely, we 
impose 
\begin{equation}
        -3\left(\frac{n}{n+1}\right)^{2}|v|^{4}
        \left|\frac{v^{2}}{g}\right|^{\frac{2}{n}} 
        + \Lambda^{4}_{SUSY} = 0.
\end{equation}
In supergravity the gravitino acquires a mass
\begin{equation}
        m_{3/2} \simeq \frac{\Lambda^{2}_{SUSY}}{\sqrt{3}}
        = \left(\frac{n}{n+1}\right) |v|^{2}
        \left|\frac{v^{2}}{g}\right|^{\frac{1}{n}}.
        \label{gravitino-mass}
\end{equation}

The inflaton $\phi$ has a mass $m_{\phi}$ in the vacuum with
\begin{equation}
        m_{\phi} \simeq n |g|^{\frac{1}{n}}|v|^{2-\frac{2}{n}}.
        \label{inftaton-mass}
\end{equation}
The inflaton $\phi$ may decay into ordinary particles through
gravitationally suppressed  interactions, which yields  reheating
temperature $T_R$ given by
\begin{equation}
    \label{reheat-temp}
    T_R \simeq 0.1 m_{\phi}^{3/2} \simeq 0.1n^{\frac{3}{2}}
    |g|^{\frac{3}{2n}}|v|^{3-\frac{3}{n}}.
\end{equation}
For example, the reheating temperature $T_R$ is as low as $2 - 6 
\times 10^{4}$GeV for $v \simeq 10^{-8} - 10^{-6}$ ($m_{3/2} \simeq 
0.02 {\rm GeV}-2$TeV), $n=4$ and $g\simeq 1$, and hence the present 
model is consistent even with the existence of light gravitino 
($m_{3/2} \lesssim 1$GeV) in the gauge-mediated SUSY-breaking model.

Let us discuss the dynamics of the new inflation. From
eq.(\ref{new-pot}) the effective potential for $\phi < \langle
\phi \rangle$
is written approximately as 
\begin{equation}
    \label{new-eff-pot}
    V \simeq |v^2 - g\phi^n|^2 - \kappa v^4 |\phi|^2.
\end{equation}
Then, identifying the inflaton field $\varphi(x)/\sqrt{2}$ with the
real part of the field $\phi(x)$, we obtain a potential for the
inflaton,
\begin{equation}
    \label{new-eff-pot2}
    V(\varphi) \simeq v^4 - \frac{\kappa}{2}v^4\varphi^2
    -\frac{g}{2^{\frac{n}{2}-1}}v^2\varphi^n 
    + \frac{g^2}{2^n}\varphi^{2n}.
\end{equation}
It has been shown in Ref.~\cite{Izawa2} that the slow-roll condition
for the inflaton is satisfied for $\kappa < 1$ and $\varphi \lesssim
\varphi_f$ where
\begin{equation}
    \label{new-inflaton-final}
    \varphi_f \simeq \sqrt{2}
    \left(\frac{(1-\kappa)v^2}{gn(n-1)}\right)^{\frac{1}{n-2}}.
\end{equation}
This provides the value of $\varphi$ at the end of inflation. The
Hubble parameter during the new inflation is given by
\begin{equation}
    \label{new-hubble}
    H \simeq \frac{v^2}{\sqrt{3}}.
\end{equation}
The scale factor of the universe increases by a factor of $e^N$ when 
the inflaton $\varphi$ rolls slowly down the potential from 
$\varphi_N$ to $\varphi_f$.  The $e$-fold number $N$ is given by 
\begin{eqnarray}
    N & \simeq & \int^{\varphi_{N}}_{\varphi_f}
    d\varphi \frac{V}{V'}
    \nonumber \\
    & = &  \left\{\begin{array}{ll}
          \frac{1}{\kappa} 
          \ln\left(\frac{\tilde{\varphi}}{\varphi_{N}}\right)
          + \frac{1-n\kappa}{(n-2)\kappa (1-\kappa)}  
          & ( \kappa \le \frac{1}{n}), \\[1em]
          \frac{1}{\kappa} 
          \ln\left(\frac{\varphi_f}{\varphi_{N}}\right)
          & ( \kappa > \frac{1}{n}),
      \end{array} \right. 
    \label{N-efold2}
\end{eqnarray}
where  
\begin{equation}
    \tilde{\varphi}  =  \sqrt{2}
    \left(\frac{\kappa v^{2}}{gn}\right)^{\frac{1}{n-2}}.
\end{equation}

The amplitude of primordial density fluctuations $\delta \rho/\rho$ 
due to the new inflation is written as 
\begin{equation}
    \label{new-density}
    \frac{\delta\rho}{\rho} \simeq \frac{1}{5\sqrt{3}\pi}
    \frac{V^{3/2}(\varphi_{N})}{|V'(\varphi_{N})|}
    = \frac{1}{5\sqrt{3}\pi} \frac{v^{2}}{\kappa\varphi_{N}}.
\end{equation}
Notice  here that we have large density fluctuations for small $\varphi_N$.
Another interesting point on the above density fluctuations is that it
results in the tilted spectrum whose spectrum index $n_s$ is given 
by~\cite{Izawa,Izawa2} 
\begin{equation}
    \label{new-index}
    n_s \simeq 1 - 2 \kappa.
\end{equation}
As is shown later, we assume $\kappa \sim 0.3$ and $n_{s}\sim 0.4$.  
This tilted power spectrum is crucial for suppressing the formation of 
small primordial black holes.  

The $e$-fold number $N$ is related to
the present cosmological scale $L$ by 
\begin{equation}
    N \simeq 60 + \ln\left(\frac{L}{3000{\rm Mpc}}\right).
    \label{N-efold}
\end{equation}
Thus, if the total $e$-fold number $N_{tot}$ of the new inflation is larger
than about 60, the new inflation accounts for all cosmological
scales of the present universe and the COBE normalization~\cite{COBE}
gives
\begin{equation}
    \frac{V^{3/2}(\varphi_{60})}{|V'(\varphi_{60})|}
    \simeq 5.3\times 10^{-4},
\end{equation}
On the other hand, if $N_{tot} < 60$, the new inflation can only 
provide density fluctuations corresponding to small scales of the 
universe and the pre-inflation discussed in the next section must 
account for density fluctuations on the large scales of 
the present universe.\footnote{
In the case of $N_{tot} \lesssim 60$, we have a domain wall problem 
since the discrete $Z_{2nR}$ symmetry is spontaneously broken. However, 
we can avoid this problem by introducing a tiny $Z_{2nR}$ 
breaking term without affecting the inflation dynamics. }

\section{A pre-inflation model}

In this section we discuss pre-inflation which occurs before the new 
inflation.  In Ref.~\cite{Izawa} it has been pointed out that the initial 
value $\varphi(x)$ required for the new inflation is dynamically tuned 
by the pre-inflation.  Here we adopt a hybrid inflation model in 
ref.~\cite{hybrid} as the pre-inflation.

The hybrid inflation model contains two kinds of superfields: one is 
$S(x,\theta)$ and the others are $\Psi(x,\theta)$ and 
$\bar{\Psi}(x,\theta)$. The model is also based on the $U(1)_R$  
symmetry. The superpotential is given by~\cite{Copeland,hybrid}\footnote{
Symmetries of this model are discussed in Ref.~\cite{Copeland,hybrid}.}
\begin{equation}
    W = -\mu^{2} S + \lambda S \bar{\Psi}\Psi.
\end{equation}
The $R$-invariant K\"ahler potential is given by
\begin{equation}
    K(S,\Psi,\bar{\Psi}) = |S|^{2} + |\Psi|^{2} + |\bar{\Psi}|^{2}
    -\frac{\zeta}{4}|S|^{4} + \cdots .
\end{equation}
The potential in supergravity is given by 
\begin{equation}
    V \simeq |\mu^{2} - \lambda\bar{\Psi}\Psi|^{2}
    + |\lambda \Psi S|^{2} + |\lambda\bar{\Psi}S|^{2}
        + \zeta\mu^{4}|S|^{2} + \cdots .
\end{equation}
The real part 
of $S(x)$ is identified with the inflaton field $\sigma/\sqrt{2}$.  
The potential is written as 
\begin{equation}
    V\simeq |\mu^{2} - \lambda \bar{\Psi}\Psi |^{2} +
    \frac{|\lambda|^{2}}{2}\sigma^{2}(|\Psi|^{2}+|\bar{\Psi}|^{2}) +
    \frac{\zeta}{2}\mu^{4}\sigma^{2} + \cdots.
\end{equation}
We readily see that if the universe starts with sufficiently large 
value of $\sigma$, the inflation occurs for $0< \zeta <1$ and 
continues until $\sigma \simeq \sigma_{c}= 
\sqrt{2}\mu/\sqrt{\lambda}$.  

In a region of small $\sigma$ 
($\sigma_{c}\lesssim \sigma \lesssim \lambda /\sqrt{8\pi^{2}\zeta}$) 
radiative corrections become important for the inflation dynamics as 
shown by Dvali {\it et al}.\cite{Dvali}.  Including one-loop 
corrections, the potential for the inflaton $\sigma$ is given by 
\begin{equation}
    \label{pre-eff-pot}
    V \simeq \mu^4 ( 1 + \frac{\zeta}{2}\sigma^{2} 
    + \frac{\lambda^2}{8\pi^2}\ln\left(\frac{\sigma}{\sigma_c}\right)).
\end{equation}
The Hubble parameter and $e$-folding factor $N'$ are given by
\begin{equation}
    H \simeq \frac{\mu^{2}}{\sqrt{3}},
\end{equation}
and
\begin{equation}
    N' \simeq   \left\{ \begin{array}{ll}
          \frac{1}{2\zeta} +\frac{1}{\zeta}
          \ln\frac{\sigma_{N'}}{\tilde{\sigma}}
          & (\sigma_{N'} >\tilde{\sigma} ),\\
          \frac{4\pi^2\sigma_{N'}^2}{\lambda^{2}}
          & (\sigma_{N'} < \tilde{\sigma} ),
      \end{array}\right.
    \label{Ndash}
\end{equation}
where 
\begin{equation}
    \tilde{\sigma} \simeq \frac{\lambda}{2\sqrt{2\zeta}\pi}.
\end{equation}

The crucial point observed in Ref.~\cite{Izawa} is that the 
pre-inflation sets dynamically the initial condition for the new 
inflation.  The inflaton field $\varphi(x)$ for the new inflation gets 
an effective mass $\sim \mu^2$ from the $e^{K}V$ term~\cite{Gaillard} 
during the pre-inflation. The precise value of the effective mass
depends on the details of the K\"ahler potential. For an example, if 
the K\"ahler potential contains $-k|\phi|^2|S|^2$, the effective mass  
is equal to $\sqrt{1+k}\mu^2$. Thus, taking account of this uncertainty
we write the effective mass $m_{eff}$ as
\begin{equation}
    \label{eff-mass}
    m_{eff} = c\mu^2 = \sqrt{3} c H,
\end{equation}
where $c$ is a free parameter. For $c \gtrsim 1$ this effective mass
is larger than the Hubble parameter for the pre-inflation. Therefore,
the $\varphi$ oscillates during the pre-inflation and its amplitude
decreases as $a^{-3/2}$ where $a$ denotes the scale factor of the
universe.  Thus, at the end of the pre-inflation the $\varphi$ takes a
value
\begin{equation}
    \varphi \simeq \varphi_{i} e^{-\frac{3}{2}N'_{tot}},
    \label{coherent}
\end{equation}
where $\varphi_{i}$ is the value of $\varphi$ at the beginning of the 
pre-inflation and $N'_{tot}$ the total $e$-fold number of the 
pre-inflation.

As pointed in Ref.~\cite{Izawa}, the minimum of the potential for
$\varphi$ deviates from zero through the effect of $|D_{S}W|^2 +
|D_{\phi}W|^2 -3|W|^2$ term (see eq.(\ref{new-pot})) .  The effective
potential for $\varphi$ during the pre-inflation is written as
\begin{equation}
    V(\varphi) \simeq \frac{1}{2} c^{2}\mu^{4}\varphi^{2}+
    \frac{\sqrt{2}}{\sqrt{\lambda}}v^{2}\mu^{3}\varphi
    + \cdots.
\end{equation}
This potential has a minimum
\begin{equation}
    \label{deviation}
    \varphi_{\rm min} \simeq -\frac{\sqrt{2}}{c^{2}\sqrt{\lambda}}
    v\left(\frac{v}{\mu}\right).
\end{equation}
This determines the mean initial value 
$\varphi_b$ of the inflaton for the new inflation as
\begin{equation}
    \label{init-new-inflaton}
    \varphi_b \simeq 
    \frac{\sqrt{2}}{c^{2}\sqrt{\lambda}}
    v\left(\frac{v}{\mu}\right).
\end{equation}

We now discuss quantum effects during the pre-inflation.  It is known
that in the de Sitter universe massless fields have quantum
fluctuations whose amplitude is given by $H/(2\pi)(H/m_{eff})^{1/2}$.
If the fluctuations of the inflaton $\varphi$ were large, we would yet
have the initial value problem.  Fortunately, the quantum fluctuations
for $\varphi$ are strongly suppressed~\cite{Enquvist} since the mass
of $\varphi$ during the pre-inflation is not less than the Hubble
parameter for $c \gtrsim 1$.  In fact, the amplitude of fluctuations
with wavelength corresponding to the horizon scale at the beginning of
the
new inflation is given by 
\begin{equation}
     \delta \varphi \simeq
     \frac{H}{2\pi}
     \left(\frac{H}{m_{eff}}\right)^{\frac{1}{2}}
     \exp[-(3/2)\ln(\mu/v)]
     \simeq \frac{H}{3^{1/4}c^{1/2}2\pi}
     \left(\frac{v}{\mu}\right)^{\frac{3}{2}}.
     \label{q-fluctuation}
\end{equation}
Here we have assumed that the reheating takes place soon after the 
pre-inflation.\footnote{
Since the reheating temperature after the pre-inflation is high, 
gravitinos are produced at the reheating epoch.  However, these 
gravitinos are sufficiently diluted by the new inflation if 
$N_{tot}\gtrsim 7$.} 
When the total $e$-fold of the new inflation is less than $60$, this 
amplitude should be much less than $\phi_b$, otherwise the present universe 
becomes very inhomogeneous.\footnote{
We thank J. Yokoyama for this point.}
Thus, we require the the ratio $R$ of $\delta\varphi$ to $\varphi_{b}$
should be much less than $1$;
\begin{equation}
    \label{init-new-constraint}
    R \equiv \frac{\delta\varphi}{\varphi_b} \ll 1.
\end{equation}

The pre-inflation also produces the density fluctuations $\delta
\rho/\rho$ with amplitude given by 
\begin{equation}
    \label{pre-density}
    \frac{\delta\rho}{\rho} \simeq \frac{1}{5\sqrt{3}\pi}
    \frac{V^{3/2}(\sigma_{N'})}{|V'(\sigma_{N'})|}
    = \left\{ \begin{array}{ll}
        \frac{1}{5\sqrt{3}\pi} \frac{\mu^{2}}{\zeta\sigma_{N'}} 
        & ( \sigma_{N'} > \tilde{\sigma} ),\\[1em]
       \frac{1}{5\sqrt{3}\pi} \frac{8\pi^2\mu^{2}\sigma_{N'}}{\lambda^2} 
        & ( \sigma_{N'} < \tilde{\sigma} ).
    \end{array}\right. 
\end{equation}
The spectral index $n_{s}$ is almost $1$ for $\sigma_{N'} < 
\tilde{\sigma}$ and $1+2\zeta$ for $\sigma_{N'} > \tilde{\sigma}$.

As mentioned in the previous section, if the new inflation provides 
sufficiently large $e$-fold number (i.e.  $N_{tot} > 60$) the density 
fluctuations produced in the pre-inflation are cosmologically 
irrelevant.  However, in the case of $N_{tot} < 60$, density 
fluctuations of the pre-inflation should account for the large scale 
structure of the universe, while the new inflation gives density 
fluctuations relevant only for formation of small scale 
structures.  In this case the amplitude of the density fluctuations 
produced in the new inflation is free from the COBE normalization and 
hence much larger density fluctuations may be produced.  Such large 
density fluctuations may yield a large number of primordial black 
holes whose mass depends on $N_{tot}$.  The most interesting case is 
that the black holes have mass $\sim 1M_{\odot}$ and explain the 
MACHOs in our galaxy.  Therefore, in the following sections, we study 
the formation of $\sim 1M_{\odot}$ black holes in our double inflation 
model.

\section{Black hole formation}

In a radiation dominated universe, primordial black holes are formed 
if the density fluctuations $\delta $ at horizon crossing satisfy a 
condition $1/3 \le \delta \le 1$~\cite{Car,Green}.  Masses of the 
black holes $M_{BH}$ are roughly equal to the horizon mass: 
\begin{equation}
    \label{bh-mass}
    M_{BH} \simeq 4\sqrt{3}\pi\frac{1}{\sqrt{\rho}}
    \simeq 0.066 M_{\odot} \left(\frac{T}{{\rm GeV}}\right)^{-2},
\end{equation}
where $\rho$ and $T$ are the total density and temperature of the
universe, respectively. Thus, the black holes with mass $\sim
1M_{\odot}$ can be formed at temperature $\sim 0.26$GeV. Since we are
interested in the black holes to be identified with the MACHOs, we assume
hereafter that the temperature of the black hole formation is
$T_{*}\simeq 0.26$GeV. 

The mass fraction $\beta_{*} (= \rho_{BH}/\rho)$ of the primordial 
black holes with mass $M_{*} \sim 1M_{\odot}$ is given by~\cite{Green} 
\begin{equation}
    \label{bh-frac}
    \beta_{*}(M_{*}) = \int_{1/3}^{1} 
    \frac{d\delta}{\sqrt{2\pi}\bar{\delta}(M_*)}
    \exp\left(-\frac{\delta^2}{2\bar{\delta}^2(M_{*})}\right)
    \simeq \bar{\delta}(M_{*}) 
    \exp\left(-\frac{1}{18\bar{\delta}^2(M_{*})}\right) ,
\end{equation}
where $\bar{\delta}(M_{*})$ is the mass variance at horizon
crossing. Assuming that only black holes with mass $M_{*}$ are formed
(this assumption is justified later), the density of the black holes
$\rho_{BH}$ is given by
\begin{equation}
    \label{bh-density}
    \frac{\rho_{BH}}{s} \simeq \frac{3}{4}\beta_{*}T_{*},
\end{equation}
where $s$ is the entropy density.  Since $\rho_{BH}/s$ is constant at 
$T<T_{*}$, we can write the density parameter $\Omega_{BH}$ of 
the black holes in the present universe as 
\begin{equation}
    \Omega_{BH}h^2 \simeq 5.6\times 10^7 \beta_{*},
\end{equation}
where we have used the present entropy density $2.9\times 
10^{3}$cm$^{-3}$ and $h$ is the present Hubble constant in units of 
100km/sec/Mpc.  Requiring that the black holes ($=$MACHOs) are dark 
matter of the universe, i.e.  $\Omega_{BH}h^2 \sim 0.25$, we obtain 
$\beta_{*} \sim 5\times 10^{-9}$ which leads to 
\begin{equation}
    \bar{\delta} (M_{*}) \simeq 0.06.
\end{equation}
This mass variance suggests that the amplitude of the density
fluctuations  at the mass scale $M_{*}$ are given by 
\begin{equation}
        \frac{\delta \rho}{\rho} \simeq 2\Phi  \simeq 0.01,
        \label{delta-rho-BH}
\end{equation}
where $\Phi$ is the gauge-invariant fluctuations of the gravitational
potential~\cite{Linde-book}.  We will show later that such
large density fluctuations are naturally produced during the new
inflation.

Since only fluctuations produced during the new inflation have
amplitudes large enough to form the primordial black holes, the
maximum mass of the black holes is determined by the fluctuations with
wavelength equal to the horizon at the beginning of the new inflation.
We require that the maximum mass is $\sim 1M_{\odot}$.  On the other
hand, the formation of black holes with smaller masses is suppressed
since the spectrum of the density fluctuations predicted by the new
inflation is tilted (see eq.(\ref{new-index})): the amplitude of the
fluctuations with smaller wavelength is smaller.  The tiny decrease
of $\bar{\delta}(M)$ results in large suppression of the black hole
formation rate as is seen from eq.(\ref{bh-frac}).  Therefore, only
black holes with mass in a narrow range are formed in the present
model.

The horizon length at the black hole formation epoch ($T=T_*$)
corresponds to scale $L_{*}$ in the present universe given by 
\begin{equation}
    \label{bh-scale}
    L_{*} \simeq \frac{a(T_0)}{a(T_{*})}H^{-1}(T_{*})
    \simeq 0.25 {\rm pc},
\end{equation}
where $T_{0}$ is the temperature of the present universe.  From 
eq.(\ref{N-efold}), the density fluctuations corresponding to $L_{*}$ 
are produced when $N = N_{*} \simeq 40$ during the new inflation.  
Since the initial value $\varphi_{b}$ for the new inflation is given 
by $\varphi_{b} =\varphi_{N_{*}}$, we obtain 
\begin{equation}
    \label{new-density2}
    \frac{V^{3/2}}{V} \simeq \frac{v^2}{\kappa \varphi_b} \simeq 
    0.3,
\end{equation}
where $\varphi_b$ is given by eq.(\ref{init-new-inflaton}) and we 
have used eq.(\ref{delta-rho-BH}).  Eqs.(\ref{new-inflaton-final}) and 
(\ref{N-efold2}) lead to 
\begin{equation}
    \label{v-k}
    v \simeq 0.3 \kappa \exp(-\kappa N_{*}),
\end{equation}
where we have taken $n=4$ and $\sqrt{(1-\kappa)/(6g)}\simeq
1$. Then the gravitino mass $m_{3/2}$ and the reheating temperature $T_{R}$
are given by
\begin{eqnarray}
    m_{3/2} & \simeq & 4.0\times 10^{-2}\kappa^{2.5}
    \exp(-2.5\kappa N_{*}),\\
    T_{R} & \simeq & 5.3 \times 10^{-2}\kappa^{2.25}
    \exp(-2.25\kappa N_{*}),\\
\end{eqnarray}
which give $m_{3/2} \simeq (0.041 -400)$GeV and $T_R \simeq (3.8 -
1.6\times 10^4)$GeV for $\kappa \simeq 0.3-0.4$.\footnote{
Since the mass of the gravitino becomes larger than $\sim 1$TeV for
$\kappa \lesssim 0.29$, we take $\kappa \gtrsim 0.29$.}
Thus the present model does not have the gravitino problem as 
mentioned before.  Using eqs.(\ref{init-new-inflaton}) and 
(\ref{new-density2}) we write the scale of the pre-inflation $\mu$ 
as 
\begin{equation}
    \label{mu-c}
    \mu \simeq 0.42 \frac{\kappa}{\lambda^{1/2}c^{2}}.
\end{equation}

The density fluctuations produced in the pre-inflation should be
normalized by the COBE data.  For this end we must take into account
the fact that the fluctuations produced at the late stage of the
pre-inflation re-enter the horizon before the beginning of the new
inflation.  Such fluctuations are cosmologically irrelevant since the
new inflation produces much larger fluctuations.  Thus, the COBE scale
corresponds to
the $e$-fold number of the pre-inflation given by 
\begin{equation}
    \label{N-COBE}
    N_{\rm COBE} = 60-N_{*} + \ln\left(\frac{\mu}{v}\right).
\end{equation}
Then, from eqs. (\ref{Ndash}) and (\ref{pre-density}), we get
\begin{equation}
    \label{mu-N-one-loop}
    \mu \simeq 6.5\times 10^{-3} \lambda^{1/2} N_{\rm COBE}^{-1/4},
\end{equation}
for $\sigma_{N_{\rm COBE}}\equiv \sigma_{\rm COBE} < \tilde{\sigma}$, and
\begin{equation}
    \label{mu-N-kahler}
    \mu \simeq 6.0\times 10^{-3} \zeta^{1/4}\lambda^{1/2}
    \exp(\zeta N_{\rm COBE}/2),
\end{equation}
for $\sigma_{\rm COBE} > \tilde{\sigma}$.

First let us consider the case where the one-loop corrections control
the pre-inflation dynamics (i.e. $\sigma_{\rm COBE} <
\tilde{\sigma}$). This corresponds to a small $\zeta$ region ($\zeta
\lesssim 0.017$). We will show that there exists a parameter region
where the black hole MACHOs are produced. For this purpose, we take
$\kappa =0.3$ as an example. Using eqs.(\ref{Ndash}), (\ref{v-k}),
(\ref{mu-c}) and (\ref{mu-N-one-loop}) we obtain
\begin{eqnarray}
    \label{v-one-loop}
     v & \simeq & 5.5\times 10^{-7},\\
     \label{mu-one-loop}
    \mu & \simeq & 2.8\times 10^{-3}\lambda^{1/2},\\
    \label{sigma-COBE-one-loop}
    \sigma_{\rm COBE} & \simeq & 0.85\lambda ,\\
    \label{c-one-loop}
    c & \simeq & 6.7 \lambda^{-1/2}.
\end{eqnarray}
Here we have neglected $\ln \lambda$ corrections.
The ratio $R$ (see eq.(\ref{init-new-constraint})) is
written as
\begin{equation}
    R \simeq 
    0.049\frac{\lambda^{1/2}\mu^{3/2}c^{3/2}}{v^{1/2}}
    \simeq 0.17 \lambda^{1/2}.
\end{equation}
We require $R \lesssim 0.03$ for the fluctuations of $\varphi$ at the
beginning of the new inflation to be negligible,\footnote{
The present universe contains $e^{60}$ regions which were horizons at
the beginning of the new inflation. For $R\lesssim 0.03$, the
probability that the $e$-fold number of a region exceeds $N_* + 1.3$
is less than $e^{-60}$. Thus, the effect of quantum fluctuations of
$\varphi$ is negligible for $R\lesssim 0.03$.}
which leads to $\lambda \lesssim 3.1\times 10^{-2}$.  The lower limit
on $\lambda$ is obtained from the condition $\sigma_c \lesssim
\sigma_{\rm COBE}$, which leads to $\lambda \gtrsim 4.6\times
10^{-3}$. Therefore, for $4.6\times 10^{-3} \lesssim \lambda \lesssim
3.1\times 10^{-2}$, our double inflation model can produce the black
holes which may be identified with MACHOs.

Next we consider the case of large $\zeta$ (i.e.  $\zeta \gtrsim
0.017$).  The value of $\zeta$ cannot be larger than about $0.2$
because the spectral index $n_{s}$ of the density fluctuations near
the COBE scale becomes $1+2\zeta$ and the COBE data~\cite{COBE} give
$n_{s} = 1.2\pm 0.3$.  From eqs.(\ref{Ndash}), (\ref{v-k}),
(\ref{mu-c}) and (\ref{mu-N-kahler}), we can determine $\mu,
\sigma_{\rm COBE}$, $c$ and $R$ for various $\zeta$'s.  The result is
shown in Table~\ref{table:kahler} for $\kappa=0.3$.  By requiring $R
\lesssim 0.03$, we get upper bounds on $\lambda$ ($\equiv \lambda_{\rm
max}$).  On the other hand, the lower bounds $\lambda_{\rm min}$ are
obtained from the condition $\tilde{\sigma} < \sigma_{\rm COBE}$.
$\lambda_{\rm max}$ and $\lambda_{\rm min}$ are also shown in
Table~\ref{table:kahler}, from which it is seen that for $\zeta
\lesssim 0.04$ there exists a consistent region ($\lambda_{\rm max}
\ge \lambda_{\rm min}$).\footnote{
The allowed range of $\zeta$ becomes narrower if we take $\kappa$
larger than 0.3.}
Since the value of allowed $\zeta$ is small, the power spectrum on the 
COBE scale is almost scale invariant ($n_{s} \simeq 1-1.1$).

\section{Conclusion}

In this paper we have studied the recently proposed double inflation 
model in supergravity and discussed the formation of primordial black 
holes with mass about $1M_{\odot}$.  We have shown that in a certain 
parameter space the primordial black holes are produced with mass 
$\sim 1 M_{\odot}$ which may be identified with MACHOs in the halo of 
our galaxy.  For successful formation of the black holes it is 
important for the inflaton $\varphi$ to have a large effective mass 
($c \gtrsim 30$) during the pre-inflation.\footnote{
In this paper we have used a large quartic coupling in the K\"ahler
potential $-k|\phi|^2|S|^2$ to produce the large effective mass for
the inflaton $\varphi$. Alternative is to introduce an extra field $X$
which gives the effective mass during the pre-inflation. For example,
consider a superpotential $W_{X} = \tilde{g}X\phi^2 + mX\bar{X}$. If
$v^2 \ll m \lesssim \mu$, $X$ may have a large value $X \sim 1$ during
the pre-inflation, which gives a large effective mass
$\tilde{g}X$. After the pre-inflation $X$ and $\bar{X}$ take vacuum
expectation values $X\simeq 0$ and $\bar{X}\simeq \tilde{g}\phi^2/m$,
and hence they never affect the dynamics of the new inflation. In this
alternative model it is sufficient to take $\tilde{g}\sim 10^{-4}$ for
our purpose.}
The allowed parameter space in the present model is very
restricted. This may be due to our specific choice of the new and
pre-inflation models. Therefore, this paper may be regarded as an
existence proof of a double inflation model which accounts for the
MACHOs as primordial black holes.  If we relax the relation between
the SUSY breaking and the new inflation scales, a wider parameter
space may be allowed.

One may consider a very steep initial power spectrum with the  
power law index $n_{s} \simeq 1.4$ in order to have sufficient 
formation of primordial black holes under the COBE normalization of 
density fluctuations.  However, models with such steep initial spectra 
overly produce black holes on smaller scales.  The existence of these 
small black holes are severely constrained from the observation of 
$\gamma$ rays.  Moreover, these models are very difficult to explain 
the large scale structure of the universe.

On the other hand, our double inflation model can naturally provide 
the power spectrum which has high amplitude and shallow slope ($n_{s} < 
1$) on small scales and low amplitude and nearly scale free spectrum 
($n_{s} \sim 1$) on large scales.  This shallow slope on small scales 
and rapid jump at the horizon scale of the turning epoch from one to 
another inflation make the mass range of primordial black holes very 
narrow.

The primordial black holes are also attractive as a source of 
gravitational waves.  If the primordial black holes dominate dark 
matter of the present universe, some of them likely form binaries.  
Such binary black holes coalesce and produce significant 
gravitational waves~\cite{Nakamura} which may be detected by future 
detectors.

\section*{Acknowledgement}
We thank J. Yokoyama for useful comments.

\begin{table}[h]
  \begin{center}
     \caption{$\mu, \sigma_{\rm COBE}$, $c$, $R$, $\lambda_{\rm max}$ and 
     $\lambda_{\rm min}$ for $\kappa=0.3$ and $\zeta=0.02, 0.04, 
     0.1, 0.2$}
     \label{table:kahler}
     \begin{tabular}{|c|cccc|}
        $\zeta$ & $0.02$  & $0.04$ &  $0.1$  & $0.2$ \\
        \hline
        $\mu$   
                & $3.0\times 10^{-3}\lambda^{1/2}$ 
                & $4.8\times 10^{-3}\lambda^{1/2}$ 
                & $1.5\times 10^{-2}\lambda^{1/2}$
                & $8.9\times 10^{-2}\lambda^{1/2}$ \\
        $\sigma_{\rm COBE}$
                & $0.86 \lambda$ 
                & $1.1 \lambda$ 
                & $4.3 \lambda$
                & $75  \lambda$ \\
        $c$     
                & $6.5\lambda^{-1/2}$ 
                & $5.1\lambda^{-1/2}$ 
                & $2.9\lambda^{-1/2}$ 
                & $1.2\lambda^{-1/2}$ \\
        $R$     
                & $0.18\lambda^{1/2}$ 
                & $0.25\lambda^{1/2}$ 
                & $0.60\lambda^{1/2}$ 
                & $2.2\lambda^{1/2}$ \\
        $\lambda_{\rm max}$     
                & $2.8\times 10^{-2}$ 
                & $1.4\times 10^{-2}$ 
                & $2.5\times 10^{-3}$ 
                & $1.8\times 10^{-4}$ \\
        $\lambda_{\rm min}$     
                & $5.3\times 10^{-3}$ 
                & $1.2\times 10^{-2}$ 
                & $6.0\times 10^{-2}$ 
                & $4.4\times 10^{-1}$ \\
    \end{tabular}
  \end{center}
\end{table}

\end{document}